# Simulation of Covid-19 epidemic evolution: are compartmental models really predictive?

Marco Paggi[1,2,*]

[1]*IMT School for Advanced Studies Lucca, Piazza San Francesco 19, 55100 Lucca, Italy*
[2]*TREE-TOWER S.R.L., c/o Polo Tecnologico Lucchese, Via della Chiesa XXXII trav. I n. 231, loc. Sorbano del Giudice, 55100 Lucca, Italy*

_______________________________________________________________________

**Abstract**

Computational models for the simulation of the severe acute respiratory syndrome coronavirus 2 (SARS-CoV-2) epidemic evolution would be extremely useful to support authorities in designing healthcare policies and lockdown measures to contain its impact on public health and economy. In Italy, the devised forecasts have been mostly based on a pure data-driven approach, by fitting and extrapolating open data on the epidemic evolution collected by the Italian Civil Protection Center. In this respect, SIR epidemiological models, which start from the description of the nonlinear interactions between population compartments, would be a much more desirable approach to understand and predict the collective emergent response. The present contribution addresses the fundamental question whether a SIR epidemiological model, suitably enriched with asymptomatic and dead individual compartments, could be able to provide reliable predictions on the epidemic evolution. To this aim, a machine learning approach based on particle swarm optimization (PSO) is proposed to automatically identify the model parameters based on a *training* set of data of progressive increasing size, considering Lombardy in Italy as a case study. The analysis of the scatter in the forecasts shows that model predictions are quite sensitive to the size of the dataset used for training, and that further data are still required to achieve convergent - and therefore reliable- predictions.

_______________________________________________________________________

1. ## Introduction

The severe acute respiratory syndrome coronavirus 2 (SARS-CoV-2), or Covid-19 for the sake of brevity, led to different healthcare policies and lockdown measures worldwide. Just to mention few instances, it is useful to recall that in the Hubei region, epicenter of the Covid-19 pandemic, extreme isolation measures were taken to rapidly reduce the spread of new infections. Lockdown measures with

---

increasing severity over time were taken also in Italy, to reduce the peak of individuals requiring intensive healthcare treatment. Tracking the movement of infected people and their contact using smartphones was considered as a possible solution to extreme social distancing in South Korea. As a different strategy, policymakers in the UK initially decided to protect only the category of old people, which experienced the higher mortality rate, allow the spread of Covid-19 in the population, aiming at reaching the so-called herd-immunity. Later, they considered the need of stricter social distancing measures as well.

During the last couple of months, valuable open data have been collected by the John Hopkins University [1] across the World and by the public authorities of each Country, making them publicly available on a daily base. In addition to the total number of infected, recovered and dead individuals reported by John Hopkins University, the Italian Civil Protection Center collected regional and national open data including also the actual number of infected people, those requiring intensive healthcare, and the number of individuals treated at home [2]. Thanks to such public resources, one has seen a proliferation of contributions in newspapers and social media from experts and non-experts, with an unprecedented mass participation to the knowledge of the evolution of Covid-19 pandemic. As a positive effect of such a knowledge share, well-known Italian scientists spotted the exponential growth at the beginning of epidemic, timely invoking urgent lockdown measures [3-5]. Later, the deviation from the exponential growth towards a logistic trend, as usually occurs for an epidemic, was highlighted [6]. Proliferation of forecasts have been noticed as well, mostly based on best-fitting of the open data using exponential or logistic regression curves, and then extrapolating trends in the future. As far as one could ascertain from newspapers [7], Regione Toscana made forecasts by multiplying the number of infected individuals by the basic reproduction number, whose value has been presumably estimated based on an exponential fit of the daily trend of the observed infectious individuals. In many other cases, analogies between Italy and Hubei, supported by the similar population size involved in the two regions, have been put forward. In general, official forecasts seem to be underestimating the actual contagion [8].

From the scientific viewpoint, all the previous attempts to forecast the evolution of Covid-19 epidemic have to be considered with extreme care, and the reliability of those predictions are highly questionable and likely valid over short periods. This is so at least for the following major reasons: (i) the emergent dynamics does depend on the lockdown measures taken over time, whose effect can be seen in the observed data only after a certain time delay; (ii) the speculative analogy between Italy and Hubei is not valid since the severity of social distancing was different in the two cases, and for the same reason each Italian Region should be analyzed separately from the others; (iii) fitting and extrapolations done on each single compartment of individuals (infectious, recovered, dead) separately from the others does not include the underlying nonlinear relations between them; (iv) since Covid-19 epidemic presents a high number of asymptomatic individuals [9], then they do affect the epidemic evolution and this information is missing in data-driven approaches; (v) logistic or exponential trends cannot actually fit the decaying trend of actually infected people after reaching the peak of infections, and therefore one could expect a lack of reliable predictions of the duration of epidemic, which is an important information to decide when removing or at least reducing lockdown.

For all of the above reasons, moving to epidemiological models could be preferable over pure data-driven approaches. Compartmental SIR (Susceptible-Infectious-Recovered compartments) models could offer a theoretical basis for the interpretation of Covid-19 epidemic. As compared to basic models with only three classes, there are variants including also dead and asymptotic individuals [10]. However, if on one side increasing the number of compartments makes the model more realistic, on the other hand the number of model parameters to be estimated substantially increases and their identification is an issue. In the case of Covid-19 epidemic, the ratio between symptomatic and asymptomatic individuals is indeed the most difficult parameter to be identified, since there are no clear data to rely on [11]. Similarly, the rate of removal of asymptomatic from infectious people after recovery is also unknown.

In this study, far from making attempts to forecast the evolution of Covid-19 epidemic, we propose a methodological contribution based on machine learning to foster the use of epidemiological models over pure data-driven best-fitting approaches and assess the reliability of their predictions. First, considering the simplified A-SIR model augmented with the asymptomatic compartment recently proposed for Covid-19 in [11], and introducing also the compartment of dead individuals, a machine learning approach based on particle swarm optimization is proposed to automatically identify all the model parameters, based on the available observations concerning actual infectious, dead and recovered individuals. Second, the problem of reliability of the A-SIR predictions is questioned and thoroughly investigated, to asses if the A-SIR model could be used to interpret the phenomenon only at the end of epidemic, or if it could be used also as a reliable forecasting tool.

## 2. The A-SIR epidemiological model with asymptomatic and dead compartments

Susceptible-Infectious-Recovered (SIR) epidemiological models [12] including three compartments have been subject of intensive research which led to several variants which included other compartments such as dead individuals (D) and asymptomatic ones (A). Such models ca be used to mathematically describe the epidemic evolution leading to a partition of a population $N$ into susceptible individuals that can become infectious over time, symptomatic or asymptomatic, up to their recovery or death. The addition of new compartments to the basic model makes it closer to reality, but at the same time it introduces additional model parameters which rises the issue of their identification.

For Covid-19 epidemic, asymptomatic infectives are considered to be a large undetected part of population [9] and, being infective for a long time not isolated, they act as a formidable vehicle for contagion. Therefore, the recent A-SIR formulation proposed in [11] is herein adopted, since it introduces this essential compartment into dynamics, albeit introducing a series of simplifying assumptions which do not increase too much model parameters. Moreover, in addition to the formulation in [11], the compartment of dead individuals is added as typically done for SIR models with vital dynamics, since daily deaths are reported by public authorities and they could be important information to test the accuracy of the model.

In the model, permanent immunity of individuals who have been infected and recovered is postulated, which is enough realistic since few cases of repeated infections have been reported so far. Susceptible individuals can evolve in one of the two classes of infected and infective people: symptomatic, *I,* and asymptomatic, *A*. People can move from the compartment of symptomatic individuals into two other compartments, one of registered recovered, *R*, and another of dead individuals, *D*. People in the compartment of asymptomatic individuals can be removed and pass to the compartment of unregistered recovered, *U*, which collects individuals passing unnoticed through the infection. In formulae, the dynamics is described by a set of nonlinearly coupled ordinary differential equations (ODEs):

$$\left(\frac{dS}{dt}\right)^{t+\Delta t} = -\beta \frac{(I+A)^t S^t}{N}, \tag{1a}$$

$$\left(\frac{dI}{dt}\right)^{t+\Delta t} = \xi\beta \frac{(I+A)^t S^t}{N} - \gamma I^t - \mu I^t, \tag{1b}$$

$$\left(\frac{dA}{dt}\right)^{t+\Delta t} = (1-\xi)\beta \frac{(I+A)^t S^t}{N} - \eta A^t, \tag{1c}$$

$$\left(\frac{dR}{dt}\right)^{t+\Delta t} = \gamma I^t, \tag{1d}$$

$$\left(\frac{dD}{dt}\right)^{t+\Delta t} = \mu I^t, \tag{1e}$$

$$\left(\frac{dU}{dt}\right)^{t+\Delta t} = \eta A^t. \tag{1f}$$

Model parameters entering Eqs. (1a)-(1f) are represented by the coefficients β, γ, μ, η, and ξ, whose identification is not an easy task, especially those related to asymptomatic individuals and unregistered recovered that have not been quantified so far. The coefficient ξ has also the physical meaning of the probability that an individual who gets infected passes to the compartment *I*, while (1–ξ) is the probability that an individual who gets infected passes to the compartment *A*.

The A-SIR differential model in Eq. (1) is coupled and nonlinear, and it can be solved by integrating the ODEs over time according to an explicit time stepping scheme (rates on the left hand side at time *t+Δt* are computed from the quantities evaluated at the previous time step, *t*), with the smallest time step possible, Δ*t*=1 day. After evaluating the rates of *S, I, A, R, U* and *D* at time *t*+Δ*t*, the updated states are computed from the previous ones as follows:

$$S^{t+\Delta t} = S^t + \left(\frac{dS}{dt}\right)^{t+\Delta t} \Delta t, \tag{2a}$$

$$I^{t+\Delta t} = I^t + \left(\frac{dI}{dt}\right)^{t+\Delta t} \Delta t, \tag{2b}$$

$$A^{t+\Delta t} = A^t + \left(\frac{dI}{dt}\right)^{t+\Delta t} \Delta t, \tag{2c}$$

$$R^{t+\Delta t} = R^t + \left(\frac{dR}{dt}\right)^{t+\Delta t} \Delta t, \tag{2d}$$

$$D^{t+\Delta t} = D^t + \left(\frac{dD}{dt}\right)^{t+\Delta t} \Delta t, \tag{2e}$$

$$U^{t+\Delta t} = U^t + \left(\frac{dU}{dt}\right)^{t+\Delta t} \Delta t. \tag{2f}$$

Initial conditions have to be specified and are given by $S^0=N$, $I^0=I_0$, $A_0=\xi/(1-\xi)I_0$, $R^0=D^0=U^0=0$. Here, $I_0$ the initial number of infected individuals reported in public open data, while $A_0$ is estimated based on ξ as suggested in [11], without introducing additional model parameters.

## 3. Model parameters' identification based on particle swarm optimization

The first step towards the application of the above A-SIR model is the identification of its 5 parameters. While the order of magnitude of γ and μ could be estimated from observations by the ratio between the daily rate of recovered or dead individuals by the actual number of symptomatic infected based on Eqs. (1d) and (1e), ξ, η and β are related to quantities that cannot be easily measured. Under such conditions, a manual identification of the values of the model parameters to match the *I, R* and *D* observations provided by the Italian Civil Protection Center is clearly difficult and might be inaccurate.

To overcome such an issue, a machine learning approach based on particle swarm optimization (PSO) is herein proposed to automatically identify the whole set of model parameters. The computational method is used to minimize the mismatch between model predictions and real observations of the individuals belonging to the *I, R* and *D* compartments by iteratively trying to improve a candidate solution. The method is therefore initialized by a population of candidate solutions in the (β, γ, μ, η, ξ)-parameter space, which are called particles, and moving these particles around in the search-space according to simple mathematical formulae over the particle's position and velocity. Each particle's movement is influenced by its local best known position, but it is also guided toward the best known positions in the search-space, which are updated as better positions are found by other particles. This meta-heuristic gradient-free method is generally expected to move the swarm toward the best solution, and it is particularly effective for nonlinear optimization problems.

The algorithm specialized to the present problem is sketched in Alg. 1.

(1) **for** each particle $x_i$, i=1,...,n, **do**
   (1.1) Initialize a group of random solutions (particles $x_i=(\beta_{i,0}, \gamma_{i,0}, \mu_{i,0}, \eta_{i,0}, \xi_{i,0}))$ in the feasible domain. They scattered over the search space (whose dimension is 5) as uniformly as possible, with a uniformly distributed random vector whose entries belong to the range from zero to $(\beta_{max}, \gamma_{max}, \mu_{max}, \eta_{max}, \xi_{max})$, i.e., $x_i$=rand(0; $x_{i,max}$).

   (1.2) Initialize the particle's best known position to its initial position $p_i \leftarrow x_i$
   (1.3) Solve ODEs and compute $S(x_i,t), I(x_i,t), R(x_i,t), D(x_i,t), A(x_i,t), U(x_i,t)$, for $t=1,...,t_{max}$
   (1.4) Compute $f(p_i)$=norm$\{I(x_i,t)–I^{exp}(t)\}$+norm$\{R(x_i,t)–R^{exp}(t)\}$+ norm$\{D(x_i,t)–D^{exp}(t)\}$
   (1.5) Find the swarm's best known position $g$ : $f(g)$=min$_i\{f(p_i)\}$
   (1.6) Initialize the particle's velocity: $v_i$=rand($-x_{i,max}$; $x_{i,max}$)

(2) **while** a termination criterion is not met **do**
   (2.1) **for** each particle i=1,...,n **do**
       (2.1.1) **for** each dimension d=1,...,5 **do**
           Peak random numbers $r_p$=rand(0,1), $r_g$=rand(0,1)
           Update particle's velocity: $v_{i,d} \leftarrow \omega\, v_{i,d}+\varphi_p\, r_p\, (p_{i,d}–x_{i,d})+ \varphi_g\, r_g\, (g_d–x_{i,d})$
       (2.1.2) Update particle's position: $x_i \leftarrow x_i + v_i$
       (2.1.3) **if** $f(x_i)<f(p_i)$ **then**
           Update the particle's best known position: $p_i \leftarrow x_i$
           **if** $f(p_i)<f(g)$ **then**
               Update the swarm's best known position: $g \leftarrow p_i$

Alg. 1: machine learning approach for A-SIR parameters' identification.

In Alg. 1, the function $f$ to be minimized was chosen as the norm of the absolute error between model predictions and observations regarding $I$, $R$ and $D$ datasets available from the Italian Civil Protection Center. The number of particles was selected equal to 1000 and the termination criterion was set in terms of maximum iterations achieved, $k_{max}$=1000. A larger population of particles or a larger number of iterations had no effect on the quality of results. Regarding the coefficients $\omega$, $\varphi_p$ and $\varphi_g$, they were set as follows: $\omega$=0.9–(0.9–0.5)$k/k_{max}$, $\varphi_p$= $\varphi_p$=0.5, where $k$ is an integer defining the iteration number, as suggested in the literature [13].

The identification procedure was applied to the following Italian Regions, selected because they presented different characteristics in terms of severity of contagion, lockdown measures, healthcare policies, initial date of outbreaks, and population size: Lombardia, Toscana, Veneto, Emilia Romagna, Lazio and Valle d'Aosta. Data from the beginning of Covid-19 epidemic in Italy (day 1 corresponds to the 24[th] of February 2020) till the 12[th] of April 2020 were considered in the dataset for PSO training. The initial outbreak date has been conventionally identified for each Region by the day such that the compartment of infectious individuals contained at least 10 people.

Overall, model predictions (shown with solid lines) are very close to real observations (dots) especially for the number of infectious (symptomatic) individuals (Fig. 1a), for all the Regions examined. This is notable since different lockdown measures occurred during the timeframe of analysis: the first lockdown began around the 21st of February 2020, covering eleven municipalities of the province of Lodi (Lombardia), that were included in a "red zone". The 8th of March, Italian Prime Minister Giuseppe Conte announced the expansion of the quarantine zone to cover much of northern Italy. With that decree, the initial "red zone" was also abolished, though the municipalities were still within the quarantined area. The locked down area, as of March 8, covered the entirety of the region of Lombardia, in addition to fourteen provinces in Piemonte, Veneto, Emilia-Romagna, and Marche. On the evening of the 9th of March, the quarantine measures were expanded to the entire country, coming into effect the next day. Conte announced on 11 March that the lockdown would be tightened, with all commercial and retail businesses except those providing essential services, like grocery stores, food stores, and pharmacies, closed down. Such lockdown measures have been prolonged so far till the 13th of May. Therefore, due to the above sequence of lockdown measures with increasing severity, the identified parameters of the A-SIR model considering the whole time frame available from the 24th of February 2020 till the 12th of April 2020 have to be considered as effective parameters, descriptive of the epidemic dynamics so far as a whole emergent phenomenon.

Regarding recovered individuals (Fig. 1b), a good matching between model predictions and real observations is noticed for all the regions, also for Valle d'Aosta which had few cases so far, which are however significative for a population which is about 100 times smaller than that of Lombardia.

As far as dead individuals are concerned (Fig. 1c), observed trends are overall well approximated, although an early deviation towards a plateau, not fully captured by the model, seems to take place for Lombardia and Valle d'Aosta.

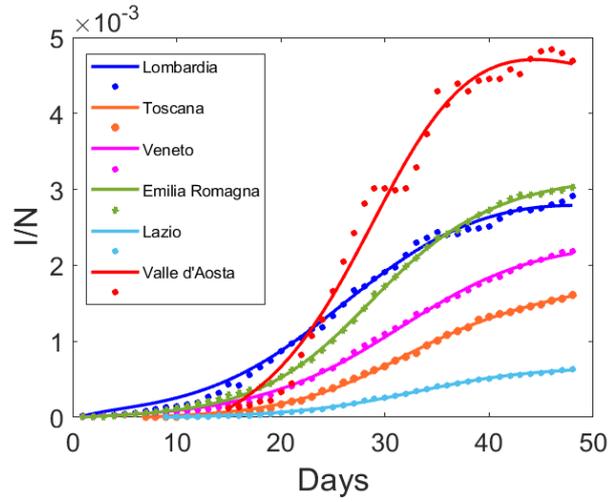

(a) Number of infectious individuals $I(t)$ divided by the population size $N$ of each Region.

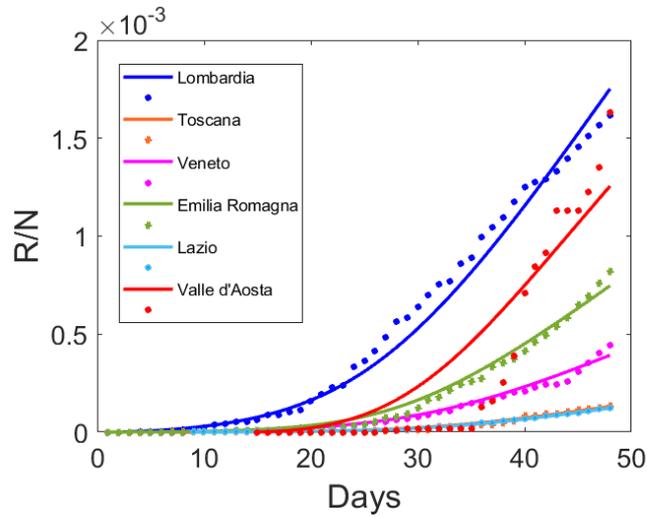

(b) Number of recovered individuals $R(t)$ divided by the population size $N$ of each Region.

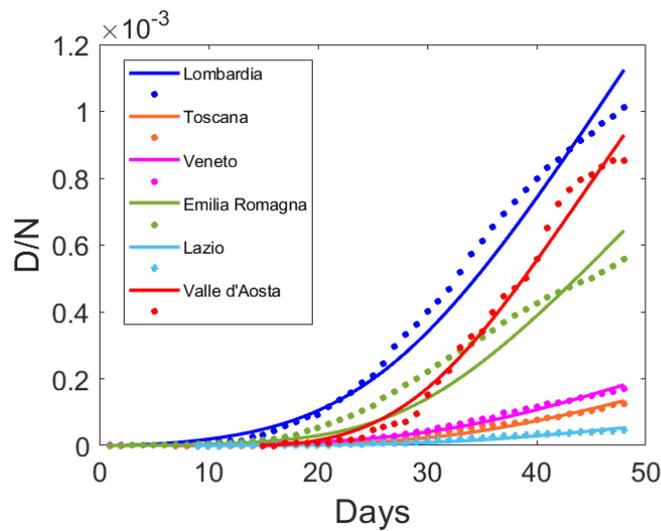

(c) Number of dead individuals $D(t)$ divided by the population size $N$ of each Region.

Fig. 1: model predictions (solid lines) vs observations (dots) for six Italian Regions.

## 4. Sensitivity of model predictions: can we use the A-SIR model as a forecasting tool?

The proposed machine learning approach for the identification of model parameters can be further exploited to test the sensitivity of model predictions based on the size of the dataset used for training. In this regard, it has to be pointed out that the PSO algorithm can be easily applied also to any other SIR model, without limitations.

Focusing on Lombardia as a case study, since it was the first Italian Region exposed to Covid-19 outbreak and therefore there are more observed data available than any other Italian Region, the PSO algorithm in Alg. 1 was applied to the dataset provided by the Italian Civil Protection Centre which reports the amount of current infectious, $I^{exp}$, recovered, $I^{exp}$, and dead, $D^{exp}$, individuals on a daily base.

The PSO algorithm was applied recursively to a dataset used for training of progressively increasing size. Its minimum size corresponds to the first 20 days of reported cases, while its maximum size is reached when all the 49 days available are included into the identification procedure. Parameters' identification was done automatically based on the training dataset, whose size is increased by adding one day per time. Once model parameters are identified, the A-SIR model is applied to forecast the evolution of epidemic up to the 150$^{th}$ day from the date of outbreak. Such predictions are shown in Fig. 2 (solid lines), while the available observed data are shown with black dots. The analysis of such predictions showed that the A-SIR model displays a significant sensitivity to the size of the dataset used for training. This is evident from the scatter in the curves that correspond to model predictions.

Overall, the A-SIR model predictions are strongly influenced by the reported actual number of infectious individuals, which is continuously increasing day after day, although its raise has been slightly slowed down during the last week. To catch up with such a raise, model predictions at the end of the epidemic have not yet reached a converged value, and therefore their reliability is questionable at the moment. This is evident from Fig. 2f, where the predicted final number of dead and recovered individuals at the end of epidemic is increasing with the day of forecast. After the 42$^{nd}$ day, such an increase is reduced, which could be an indicator of approaching the peak in the actual infectious individuals, which could lead to a convergence in the model predictions. Therefore, further observed data are still required to draw reliable conclusions and judge about the reliability of model forecasts based on the present A-SIR epidemiological model.

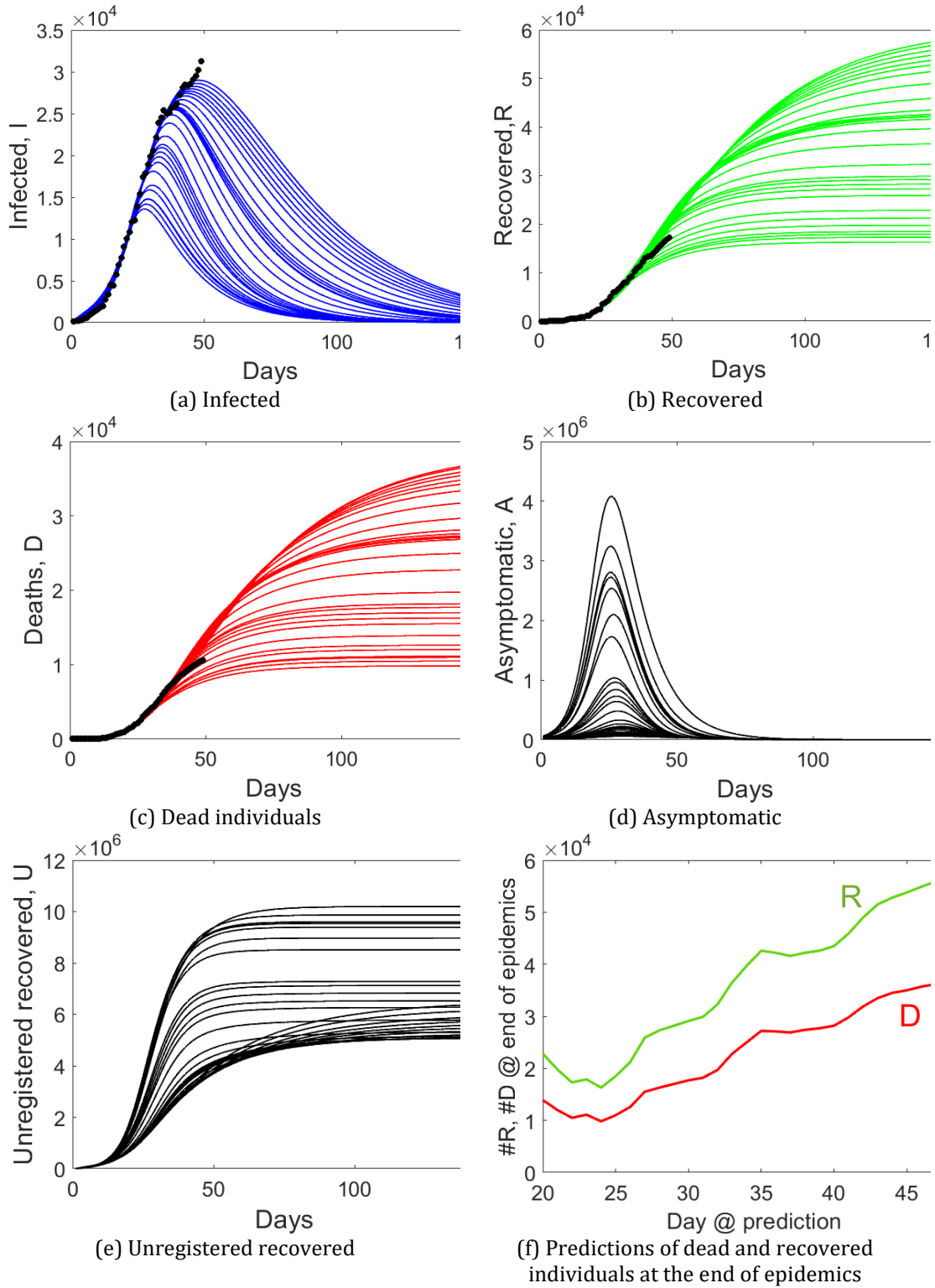

Fig. 2: scatter in the model predictions depending on the size of the dataset used for training. Its size has been varied by including an amount of days from the beginning of Covid-19 outreach in Lombardia ranging from 20 up to 49.

## 5. Discussion and preliminary conclusions

A machine learning approach of general validity has been proposed in this work to provide an automatic method to identify all the parameters of a recent A-SIR model for Covid-19 proposed in [11] and herein further extended by including also the compartment of dead individuals. The advantage of the method is that it can also identify parameters related to asymptomatic and unregistered recovered compartments, for which there are no certain observed data to rely on. Moreover, the approach can be in principle particularized to deal with any other SIR model.

The application of the PSO algorithm to a set of 6 Italian Regions has proved that the A-SIR model can effectively capture the observed trends in terms of actual infectious, recovered and dead individuals, regardless of the size of the Region and smearing out the effect of a series of different lockdown measures adopted over time though simple effective time-independent model parameters.

After proving the ability of the A-SIR model to *interpret* the already observed epidemic phenomenon, the reliability of its forecasts has been investigated by varying the size of the dataset used for model training. Results show that, so far, model predictions are still quite scattered and have not yet reached convergence. Therefore, the A-SIR model should be considered with care for long-range forecasts. Additional observations are therefore deemed to be necessary to judge its predictions' reliability.